\documentstyle[sprocl]{article}
\input psfig.sty

\def\etal{{\it et al.}}
\newcommand{\newc}{\newcommand}

\def\mw{M_W}
\def\bsg{\ifmmode B\to X_s\gamma\else $B\to X_s\gamma$\fi}
\def\bsll{\ifmmode B\to X_s\ell^+\ell^-\else $B\to X_s\ell^+\ell^-$\fi}
\def\bstt{\ifmmode B\to X_s\tau^+\tau^-\else $B\to X_s\tau^+\tau^-$\fi}
\def\shat{\ifmmode \hat{s}\else $\hat{s}$\fi}
\newc{\mHpm}{m_{H^\pm}}
\newc{\gsim}{\lower.7ex\hbox{$\;\stackrel{\textstyle>}{\sim}\;$}}
\newc{\lsim}{\lower.7ex\hbox{$\;\stackrel{\textstyle<}{\sim}\;$}}
\newc{\ie}{{\it i.e.}}          
\newc{\eg}{{\it e.g.}}          
\newc{\kev}{\hbox{\rm\,keV}}            
\newc{\mev}{\hbox{\rm\,MeV}}            
\newc{\gev}{\hbox{\rm\,GeV}}            
\newc{\tev}{\hbox{\rm\,TeV}}
\newc{\xpb}{\hbox{\rm\, pb}}
\newc{\xfb}{\hbox{\rm\, fb}}

\newc{\tbeta}{\tan\beta}
\newc{\uL}{{\tilde u_L}}
\newc{\uR}{{\tilde u_R}}
\newc{\cL}{{\tilde c_L}}
\newc{\cR}{{\tilde c_R}}
\newc{\tL}{{\tilde t_L}}
\newc{\tR}{{\tilde t_R}}
\newc{\dL}{{\tilde d_L}}
\newc{\dR}{{\tilde d_R}}
\newc{\sL}{{\tilde s_L}}
\newc{\sR}{{\tilde s_R}}
\newc{\bL}{{\tilde b_L}}
\newc{\bR}{{\tilde b_R}}
\newc{\eL}{{\tilde e_L}}
\newc{\eR}{{\tilde e_R}}
\newc{\mhp}{m_{H^\pm}}
\newc{\mhalf}{m_{1/2}}
\newc{\lR}{\tilde{l}_R}
\newc{\lL}{\tilde{l}_L}
\newc{\nL}{\tilde{\nu}_L}
\newc{\na}{\chi^0_1}
\newc{\nb}{\chi^0_2}
\newc{\nc}{\chi^0_3}
\newc{\nd}{\chi^0_4}
\newc{\ca}{\chi^{\pm}_1}
\newc{\cb}{\chi^{\pm}_2}
\newc{\camp}{\chi^\mp_1}
\newc{\cbmp}{\chi^\mp_1}
\newc{\capos}{\chi^{+}_1}
\newc{\caneg}{\chi^{-}_1}
\def\beq{\begin{equation}}
\def\eeq{\end{equation}}
\def\bea{\begin{eqnarray*}}
\def\eea{\end{eqnarray*}}
\def\slashchar#1{\setbox0=\hbox{$#1$}           
   \dimen0=\wd0                                 
   \setbox1=\hbox{/} \dimen1=\wd1               
   \ifdim\dimen0>\dimen1                        
      \rlap{\hbox to \dimen0{\hfil/\hfil}}      
      #1                                        
   \else                                        
      \rlap{\hbox to \dimen1{\hfil$#1$\hfil}}   
      /                                         
   \fi}                                         %

\begin{document}
\bibliographystyle{unsrt}    

%
%
\def\MPL #1 #2 #3 {{\em Mod. Phys. Lett.} {\bf#1},\ #2 (#3)}
\def\NPB #1 #2 #3 {{\em Nucl. Phys.} {\bf#1},\ #2 (#3)}
\def\PLB #1 #2 #3 {{\em Phys. Lett.} {\bf#1},\ #2 (#3)}
\def\PR #1 #2 #3 {{\em Phys. Rep.} {\bf#1},\ #2 (#3)}
\def\PRD #1 #2 #3 {{\em Phys. Rev.} {\bf#1},\ #2 (#3)}
\def\PRL #1 #2 #3 {{\em Phys. Rev. Lett.} {\bf#1},\ #2 (#3)}
\def\RMP #1 #2 #3 {{\em Rev. Mod. Phys.} {\bf#1},\ #2 (#3)}
\def\ZPC #1 #2 #3 {{\em Z. Phys.} {\bf#1},\ #2 (#3)}

\renewcommand{\thefootnote}{\fnsymbol{footnote}} 
\def\srf#1{$^{#1}$\ }
\def\mainhead#1{\setcounter{equation}{0}\addtocounter{section}{1}
  \vbox{\begin{center}\large\bf #1\end{center}}\nobreak\par}
\def\subhead#1{\vbox{\smallskip\noindent \bf #1}\nobreak\par}
\def\autolabel#1{\auto\label{#1}}
\hfuzz=1pt
\renewcommand{\arraystretch}{1.5}

\begin{titlepage} 
\rightline{\vbox{\halign{&#\hfil\cr
&SLAC-PUB-7774\cr
&December 1997\cr}}}
\vspace{0.5in} 
\begin{center}

{\Large\bf
B PHYSICS BEYOND THE STANDARD MODEL}
\footnote{Work supported by the Department of 
Energy, Contract DE-AC03-76SF00515}
\medskip

\normalsize 
{\large JoAnne L. Hewett \\
\vskip .3cm
Stanford Linear Accelerator Center \\
Stanford CA 94309, USA}
\vskip .3cm

\end{center}

\begin{abstract} 

The ability of present and future experiments to test the Standard Model
in the $B$ meson sector is described.  We examine the loop
effects of new interactions in flavor changing neutral current $B$ decays
and in $Z\to b\bar b$, concentrating on supersymmetry and the left-right
symmetric model as specific examples of new physics scenarios.  The procedure 
for performing a global fit to the Wilson coefficients 
which describe $b\to s$ transitions is outlined, and the results of such
a fit from Monte Carlo generated data is compared to the predictions of our 
two sample new physics scenarios.
A fit to the $Zb\bar b$ couplings from present data is also given.

\end{abstract} 

\vskip1.0in
\noindent{Based on presentations given at the {\it Workshop on Physics Beyond
the Standard Model: Beyond the Desert: Accelerator and Nonaccelerator
Approaches}, Tegernsee, Germany, June 8-14, 1997; {\it 20th Anniversary
Symposium: Twenty Beautiful Years of Bottom Physics}, Chicago, IL, June 29 -
July 2, 1997; and the {\it 7th International Symposium on Heavy Flavor
Physics}, Santa Barbara, CA, July 7-11, 1997.}

\renewcommand{\thefootnote}{\arabic{footnote}} \end{titlepage} 


\title{B PHYSICS BEYOND THE STANDARD MODEL}

\author{ J.L. HEWETT}

\address{Stanford Linear Accelerator Center, 
Stanford CA 94309, USA}

\maketitle\abstracts{
The ability of present and future experiments to test the Standard Model
in the $B$ meson sector is described.  We examine the loop
effects of new interactions in flavor changing neutral current $B$ decays
and in $Z\to b\bar b$, concentrating on supersymmetry and the left-right
symmetric model as specific examples of new physics scenarios.  The procedure 
for performing a global fit to the Wilson coefficients 
which describe $b\to s$ transitions is outlined, and the results of such
a fit from Monte Carlo generated data is compared to the predictions of our 
two sample new physics scenarios.
A fit to the $Zb\bar b$ couplings from present data is also given.}
  
\section{Overview}

The B-meson system promises to yield a fertile testing ground of the Standard
Model (SM).  The large data samples which will be acquired over the next
decade at CESR, the Tevatron, HERA, the SLAC and KEK B-factories, as well as
the LHC will furnish the means to probe the SM at an unprecedented level of
precision.  It is well-known\cite{htt} that precision measurements of 
low-energy processes can provide an insight to very high energy scales via
the indirect effects of new interactions.
As such, the $B$ sector offers a complementary probe of new physics,
and in some cases may yield constraints which surpass those from direct 
collider searches or exclude entire classes of models.

New physics may manifest itself in the B system in several ways,
for example, inconsistencies with the SM may be found in measurements of
(i) the unitarity triangle, (ii) rare decays, or (iii) precision electroweak 
measurements of the decay $Z\to b\bar b$.  
In the first case, the angles of the unitarity
triangle, commonly denoted as $\alpha\,, \beta,$ and $\gamma$, may reveal the
existence of new physics in three distinct manners:
\begin{itemize}
\item $\alpha+\beta+\gamma\neq\pi$,
\item $\alpha+\beta+\gamma = \pi$, but the individual values of the angles are
outside of their SM ranges,
\item $\alpha+\beta+\gamma = \pi$, but the values of the angles are
inconsistent with the measured sides of the triangle.
\end{itemize}
These potential deviations may originate from new interactions in tree-level 
$B$ decays, or by the virtual effects of new physics in loop mediated 
processes (\eg, $B^0_d-\bar B^0_d$ mixing or penguin decays of the $B$), with 
or without the presence of new phases.  Since the scale of the new physics is 
expected to be large compared to $M_W$, it is anticipated that additional 
tree-level contributions to $B$ decay
are suppressed.  Further consequences of new degrees of freedom in the
unitarity triangle are discussed by Fleischer.\cite{rf} 

Here, we concentrate on the loop effects of new interactions in flavor
changing neutral current (FCNC) $B$ decays and in $Z\to b\bar b$.
We note that most classes of models which induce large effects in the FCNC
decays also affect $B^0_d-\bar B^0_d$ mixing, and that measurements of several
different rare decays may elucidate the origin of new interactions.
$b\to s$ transitions provide an excellent probe of new
indirect effects as they only occur at loop level in the SM, and they have
relatively large rates for loop processes due to the massive internal-top quark
and the Cabbibo-Kobayashi-Maskawa (CKM) structure of the contributing penguin
and box diagrams.  Also, long distance effects are expected to
play less of a role due to the heavy $B$ mass, and hence rare processes
are essentially short distance dominated.

\section{The $Zb\bar b$ Vertex}

\begin{figure}[t]
\centerline{
\psfig{figure=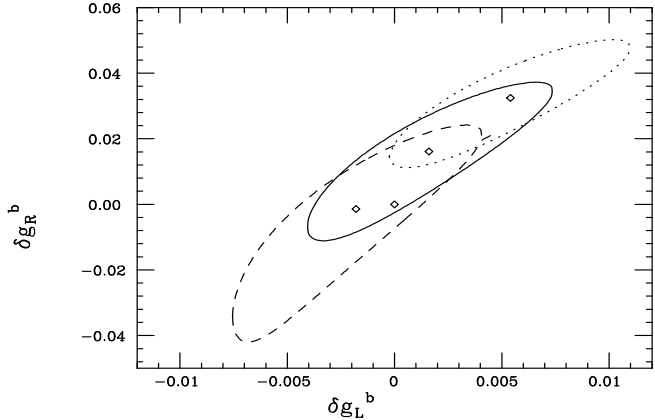,height=2.7in,width=3.95in,angle=-90}}
\vspace*{-0.50cm}
\caption{$95\%$ C.L. fit to the parameters $\delta g^b_{L,R}$ using the full
SLC/LEP $Z\to b\bar b$ data set and ZFITTER4.9 with $m_t=175$ GeV.  The
dashed, solid, and dotted curves correspond to Higgs masses of 1 TeV, 300 GeV,
and 65 GeV, respectively.  The diamond at the center represents the SM, while
the three other diamonds correspond to the location of the $\chi^2$ minima
for each Higgs mass.  The values of the other input parameters are given in
the text.}
\label{bbar}
\end{figure}

The SM continues to provide an excellent description of precision electroweak
data,\cite{precew} where the few (small) deviations may be attributed to normal
statistical fluctuations and not neccessarily to new physics.  In particular,
the observables which characterize the $Zb\bar b$ couplings, the ratio
$R_b=\Gamma(Z\to b\bar b)/\Gamma(Z\to\mbox{hadrons})$ and the forward-backward
asymmetry parameter $A_b$, are now measured to be only $\sim (1.5-2)\sigma$ away
from their SM expectations.\cite{precew,stevew}  This is in contrast to only 
a couple of years ago when $R_b$ was measured to be anomalously high, 
yielding hopeful indications of physics beyond the SM.\cite{hagiw}  In fact, 
the $Zb\bar b$ vertex has long been recognized as being sensitive to new
physics which may not affect the light fermion vertices, and now 
constrains the parameter space of some models.\cite{htt}  A model independent
fit to possible shifts in the left- and right-handed $Zb\bar b$ couplings
is presented in Fig. \ref{bbar}.  Writing these couplings as
\begin{eqnarray}
g_L^b & = & -{1\over 2}+{1\over 3}\sin^2\theta^b_w+\delta g_L^b \,, \\
g_R^b & = & {1\over 3}\sin^2\theta^b_w+\delta g_R^b \,, \nonumber
\end{eqnarray}
where $\delta g^b_{L,R}$ represents the coupling shifts from their SM values,
and $\sin^2\theta^b_w$ is the b-quark effective weak mixing angle, we use
ZFITTER4.9 to calculate the SM predictions (taking $m_t=175$ GeV, $\alpha_s
=0.118$, and $\alpha_{em}^{-1}=128.896$) and perform
a fit to the full SLC/LEP $Z\to b\bar b$ data set.  Here we see that the data
is well described by the SM, with a slight preference for a heavier Higgs.

\section{Formalism for $b\to s$ Transitions}

The observation\cite{cleo:94} of radiative penguin mediated processes, in both 
the exclusive $B\to K^*\gamma$ and inclusive \bsg\ channels, 
has placed the study of rare $B$ decays on a new footing and has provided
powerful constraints on classes of models\cite{topten}.

The effective field theory for $b\to s$ transitions is summarized at this
meeting by Ali\cite{alihf}, however, we briefly review the features which are
essential to the remainder of this talk.  Incorporating the QCD corrections, 
these transitions are governed by the Hamiltonian 
\begin{equation}
{\cal H}_{eff}={-4G_F\over\sqrt 2}V_{tb}V^*_{ts}\sum_{i=1}^{10}C_i(\mu)
{\cal O}_i(\mu)\,,
\label{effham}
\end{equation}
where the ${\cal O}_i$ are a complete set of renormalized operators of
dimension six or less which mediate these transitions and are 
catalogued in, \eg , Ref. \cite{buras:95}.  The $C_i$ represent
the corresponding Wilson coefficients which are evaluated perturbatively
at the electroweak scale, where the matching conditions are imposed, and
then evolved down to the renormalization scale $\mu\approx  m_b$.  The
expressions for $C_i(M_W)$ in the SM are given by the Inami-Lim 
functions \cite{inami}.

\subsection{$B\to X_s\ell^+\ell^-$ in the Standard Model}

For \bsll\ this formalism leads to the physical decay amplitude (neglecting 
$m_s$ and $m_\ell$)
\begin{eqnarray}
{\cal M} &=& {\sqrt 2G_F\alpha\over \pi}V_{tb}V^*_{ts}
\left[ C_9^{e\!f\!f}\bar s_L\gamma_\mu b_L\bar\ell\gamma^\mu\ell
+C_{10}\bar s_L\gamma_\mu b_L\bar\ell\gamma^\mu\gamma_5\ell  \right. 
\nonumber\\
& & \quad\left.
-2C^{e\!f\!f}_7m_b\bar s_Li
\sigma_{\mu\nu}{q^\nu\over q^2}b_R\bar\ell\gamma^\mu\ell
\right] \,, 
\end{eqnarray}
where $q^2$ represents the momentum transferred to the lepton pair.  
The next-to-leading order (NLO) analysis
for this decay has been performed in Buras \etal~\cite{buras:95}, where
it is stressed that a scheme independent result can only be obtained by 
including the leading and next-to-leading logarithmic corrections to
$C_9(\mu)$ while retaining
only the leading logarithms in the remaining Wilson coefficients.  The leading 
residual scale dependence in $C_9(\mu)$ is cancelled by that
contained in the matrix element of ${\cal O}_9$, yielding an effective value 
$C^{eff}_9$.  In addition,
the effective value for $C^{eff}_7(\mu )$ refers to the leading order 
scheme independent result, and we note that 
the operator ${\cal O}_{10}$ does not renormalize.
The numerical estimates (in the naive dimensional regularization 
scheme) for these
coefficients are displayed in Table \ref{coeffval}.
The reduced scale dependence of the NLO- versus the LO-corrected coefficients
is reflected in the deviations $\Delta C_9(\mu)\lsim\pm 10\%$ and 
$\Delta C^{eff}_7(\mu)\approx\pm 20\%$ as $\mu$ is varied.
We find that the coefficients are much less
sensitive to the remaining input parameters, with $\Delta
C_9(\mu),\Delta C^{eff}_7(\mu)\lsim 3\%$, varying 
$\alpha_s(M_Z)=0.118\pm 0.003$ \cite{pdg}, 
and $m_t^{phys}=175\pm 6\gev$ \cite{cdfd0}.
The resulting inclusive branching fractions (which are computed by scaling
the width for \bsll\ to that for $B$ semi-leptonic decay)  are found
to be $(6.25^{+1.04}_{-0.93})\times 10^{-6}$,
$(5.73^{+0.75}_{-0.78})\times 10^{-6}$, 
and $(3.24^{+0.44}_{-0.54})\times 10^{-7}$
for $\ell=e, \mu$, and $\tau$, respectively, taking into account the
above input parameter ranges, as well as 
$B_{sl}\equiv B(B\to X\ell\nu)=(10.23\pm 0.39)\%$~\cite{richman}, and
$m_c/m_b=0.29\pm 0.02$.

\begin{table}
\centering
\begin{tabular}{|c|c|c|c|} \hline\hline
Coefficient & $\mu=m_b/2$ & $\mu=m_b$ & $\mu=2m_b$ \\ \hline
$C^{eff}_7$  & $-$0.371 & $-$0.312 & $-$0.278 \\
$C_9$ & 4.52 & 4.21 & 3.81 \\
$C_{10}$ & $-$4.55 & $-$4.55 & $-$4.55 \\ \hline\hline
\end{tabular}
\caption{Values of the Wilson coefficients for several choices of the
renormalization scale.  Here, we take $m_b=4.87$ GeV, 
$m_t=175$ GeV, and $\alpha_s (M_Z)=0.118$.}
\label{coeffval}
\end{table}

\subsection{$B\to X_s\gamma$ in the Standard Model}

The basis for the decay \bsg\ contains the first eight operators
in the effective Hamiltonian of Eq.~(\ref{effham}).  The 
next-to-leading order logarithmic 
QCD corrections have been recently completed, leading to a much reduced
renormalization scale dependence in the branching fraction!
The higher-order calculation involves several steps, requiring
corrections to both $C_7$ and the matrix element of ${\cal O}_7$.
For the matrix element, this includes the QCD bremsstrahlung 
corrections \cite{greub:91} $b\to s\gamma+g$, and the NLO virtual corrections
\cite{greub:96}.  Summing these contributions to the matrix elements
and expanding them around $\mu=m_b$, one arrives at the decay 
amplitude
\begin{equation}
{\cal M}(b\to s\gamma) = -{4G_FV_{tb}V^*_{ts}\over\sqrt 2}D\langle
s\gamma|{\cal O}_7(m_b)|b\rangle_{tree} \,, 
\end{equation}
with
\begin{equation}
\label{dc7eq}
D=C_7(\mu)+{\alpha_s(m_b)\over 4\pi}\left( C_i^{(0)eff}(\mu)
\gamma_{i7}^{(0)}\log {m_b\over\mu}
+C_i^{(0)eff}r_i\right) \,.
\end{equation}
Here, the quantities $\gamma_{i7}^{(0)}$ are the entries of the effective
leading order anomalous dimension matrix, and the $r_i$ are computed in Greub
\etal~\cite{greub:96}, for $i=2,7,8$.  The first term in Eq.~(\ref{dc7eq}), 
$C_7(\mu)$, must be computed
at NLO precision, while it is consistent to use the leading order values
of the other coefficients.  
For $C_7$ the NLO result entails
the computation of the ${\cal O}(\alpha_s)$ terms in the matching conditions
\cite{yao},
and the renormalization group evolution of $C_7(\mu)$ must be computed
using the ${\cal O}(\alpha_s^2)$ anomalous dimension matrix \cite{misiak:96}.
The numerical value of the branching fraction is then found to be
(again, scaling to semi-leptonic decay)
\begin{equation}
B(\bsg)=(3.25\pm 0.30 \pm 0.40)\times 10^{-4} \,,
\end{equation}
where the first error corresponds to the combined uncertainty associated with 
the value of $m_t$ and $\mu$, and the second error represents the uncertainty
from $\alpha_s(M_Z), B_{sl}$, and $m_c/m_b$.
This is well within the range observed by CLEO\cite{cleo:94} which is
$B=(2.32\pm 0.57\pm 0.35)\times 10^{-4}$ with the $95\%$ C.L.
bounds of $1\times 10^{-4}< B(\bsg)<4.2\times 10^{-4}$.  We note that ALEPH
has recently reported the preliminary observation of this inclusive decay,
at a compatible rate \cite{aleph}.

\section{Model Independent Tests for New Physics in $b\to s$ Transitions}

\begin{figure}[t]
\centerline{
\psfig{figure=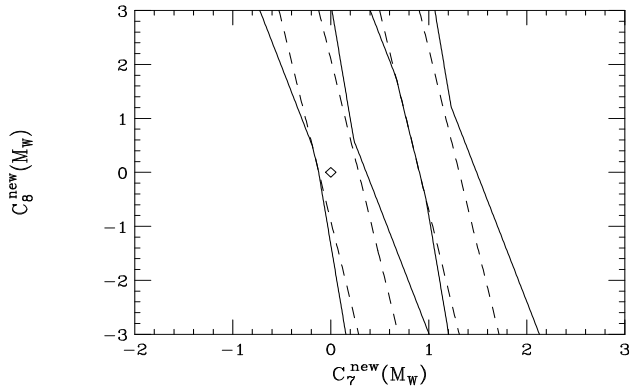,height=2.0in,width=3.25in}}
\vspace*{-0.25cm}
\caption{Bounds 
on the contributions from new physics to $C_{7,8}$.  The region allowed
by the CLEO data corresponds to the area inside the solid diagonal bands.
The dashed bands represent the constraints when the renormalization scale
is set to $\mu =m_b$. The diamond at the position
(0,0) represents the Standard Model.}
\label{fig2}
\end{figure}

Measurements of $B(\bsg)$ constrain the magnitude, but not the sign,
of $C_7(\mu)$.  The coefficients at the matching scale ($\mu=M_W$) can be 
written in the form $C_i(M_W)=C_i^{SM}(M_W)+C_i^{new}(M_W)$,
where $C_i^{new}(M_W)$ represents the contributions from new
interactions.  Due to operator mixing, \bsg\ then limits the 
possible values for $C_i^{new}(M_W)$ for $i=7,8$.  These bounds are
summarized in Fig.~\ref{fig2}.
Here, the solid bands correspond to the
constraints obtained from the current CLEO measurement,
taking into account the variation of the renormalization scale
$m_b/2 \leq \mu \leq 2m_b$, as well as the allowed ranges of
the other input parameters.  The dashed bands represent the constraints
when the scale is fixed to $\mu =m_b$.  We note that large values of
$C_8^{new}(\mw)$ (which would yield an anomalous rate for $b\to sg$)
are allowed even in the region where $C_7^{new}(\mw)\simeq 0$.  

\begin{figure}[t]
\centerline{
\psfig{figure=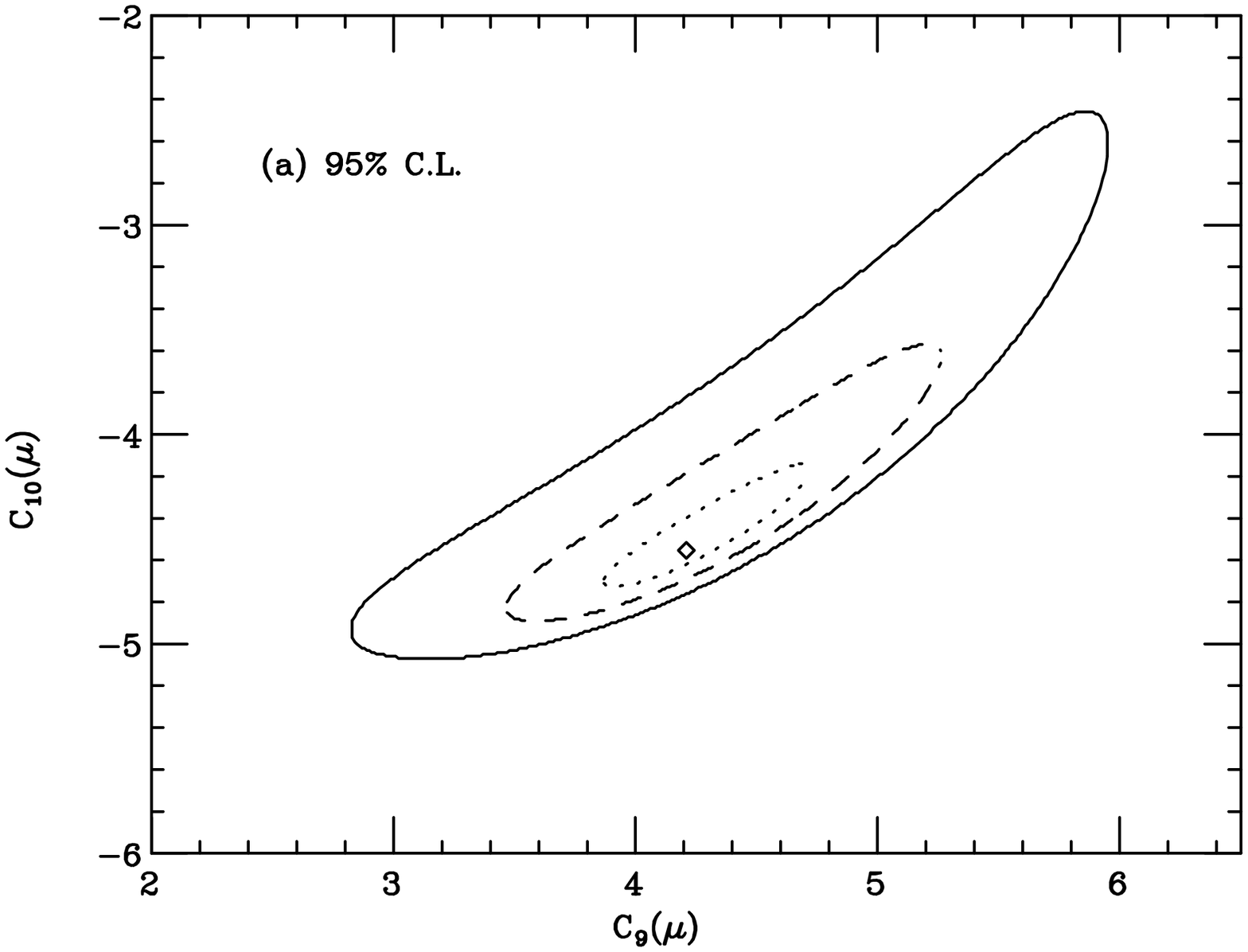,height=2.5in,width=2.50in,angle=-90}
\psfig{figure=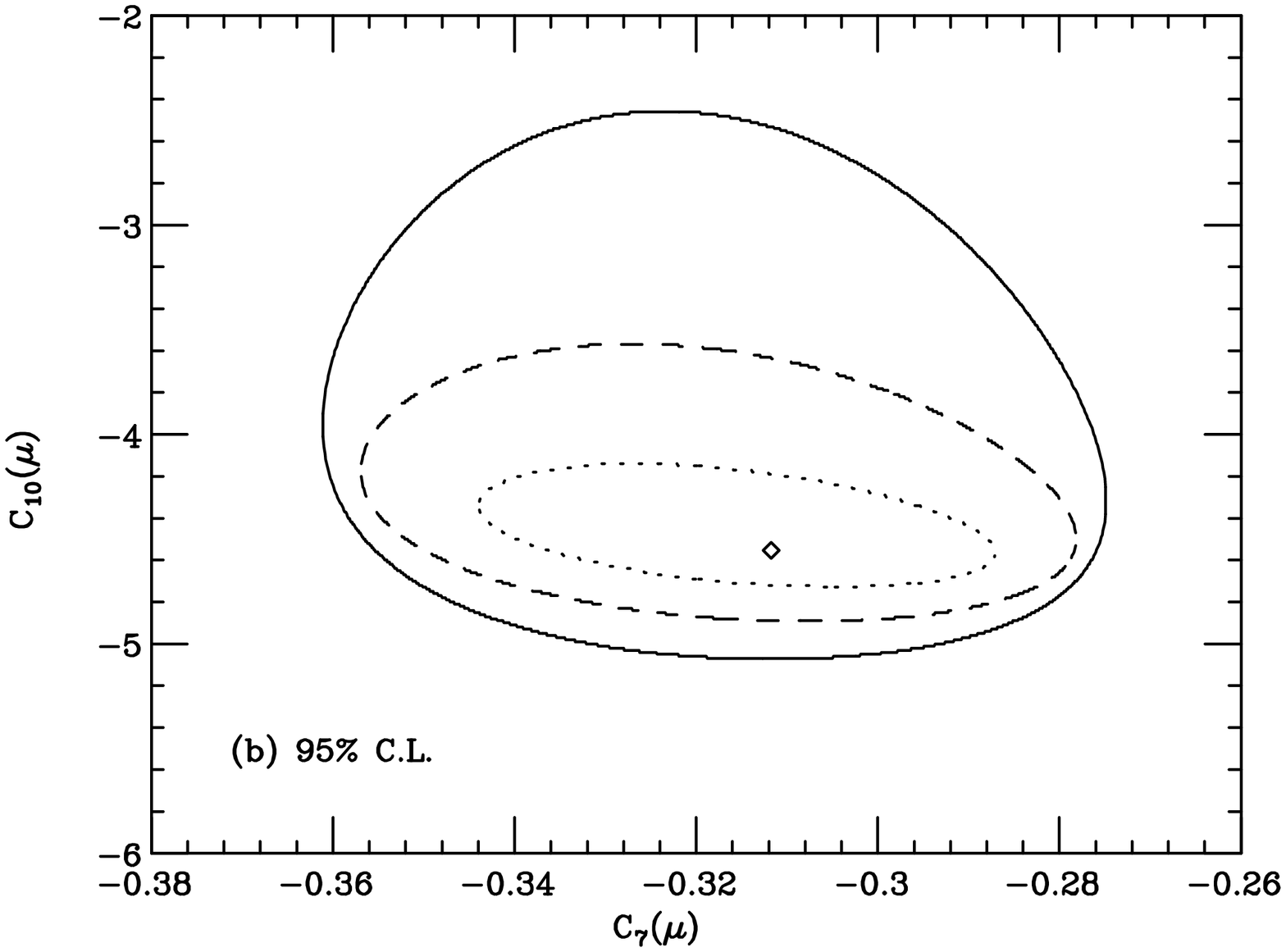,height=2.5in,width=2.50in,angle=-90}}
\vspace*{-0.25cm}
\caption{The $95\%$
C.L. projections in the (a) $C_9 - C_{10}$ and (b) $C_7 - C_{10}$ planes, 
where the allowed regions lie inside of the contours.  The solid, dashed, and 
dotted contours correspond to $3\times 10^7$, $10^8$, and $5\times 10^8$
$B\bar B$ pairs. The central value of the 
SM prediction is labeled by the diamond.}
\label{fig3}
\end{figure}

Measurement of the kinematic distributions\cite{ali,hewett} associated with the 
final state lepton pair in \bsll\ as well as the rate for \bsg\ allows 
the determination of the sign and magnitude of all the Wilson coefficients
for the contributing operators in a model independent
fashion.  We have performed a Monte Carlo analysis in order to
ascertain how much quantitative information will be obtainable at
future $B$-Factories and follow the procedure outlined in Ref. \cite{jimjo}.
For the
process \bsll, we consider the lepton pair invariant mass distribution
and forward-backward asymmetry\cite{ali} for $\ell=e, \mu, \tau$, and the tau
polarization asymmetry\cite{hewett} for \bstt.  A 
three dimensional $\chi^2$ fit to the coefficients $C_{7,9,10}(\mu)$ is 
performed for three values of integrated luminosity, 
$3\times 10^7$, $10^8$, and $5\times 10^8$ $B\bar B$ pairs,
corresponding to the expected $e^+e^-$ $B$-Factory luminosities of one
year at design, one year at an upgraded accelerator, and the total
accumulated luminosity at the end of these programs.  
The $95\%$ C.L. allowed regions (including statistical errors only for \bsll\ 
and a flat $10\%$ error on \bsg) as projected
onto the $C_9(\mu)-C_{10}(\mu)$ and $C_7(\mu)-C_{10}(\mu)$ planes are
depicted in Figs.~\ref{fig3}(a-b), where the diamond
represents the central value for the expectations in the SM given in Table 1.
We see that the determinations are
relatively poor for $3\times 10^{7}$ $B\bar B$ pairs and that
higher statistics are
required in order to focus on regions centered around the SM.  

\section{Supersymmetric Effects in $b\to s$ Transitions}

These model independent bounds can be compared
with model dependent predictions for the Wilson coefficients in order to
ascertain at what level specific new interactions can be probed.  First,
we consider supersymmetric extensions to
the SM. 
Supersymmetry (SUSY) contains many potential sources for flavor violation.  
For example, the
flavor mixing angles among the squarks are {\it a priori} 
separate from the CKM angles of the SM quarks.  
We adopt the viewpoint here that flavor-blind (diagonal) soft terms
at the high scale are the phenomenological
source for the soft scalar masses, and that the CKM
angles are the only relevant flavor violating sources.  The spectroscopy of 
the supersymmetric states is quite model dependent and we 
analyze two possibilities.  The first is the familiar
minimal supergravity model; in this instance all the
supersymmetric states follow from a common scalar mass 
and a common gaugino mass at the high scale.  The second case is where
the condition of common scalar masses is relaxed and they are allowed
to take on uncorrelated values at the low scale
while still preserving gauge invariance.

\begin{figure}[t]
\centerline{
\psfig{figure=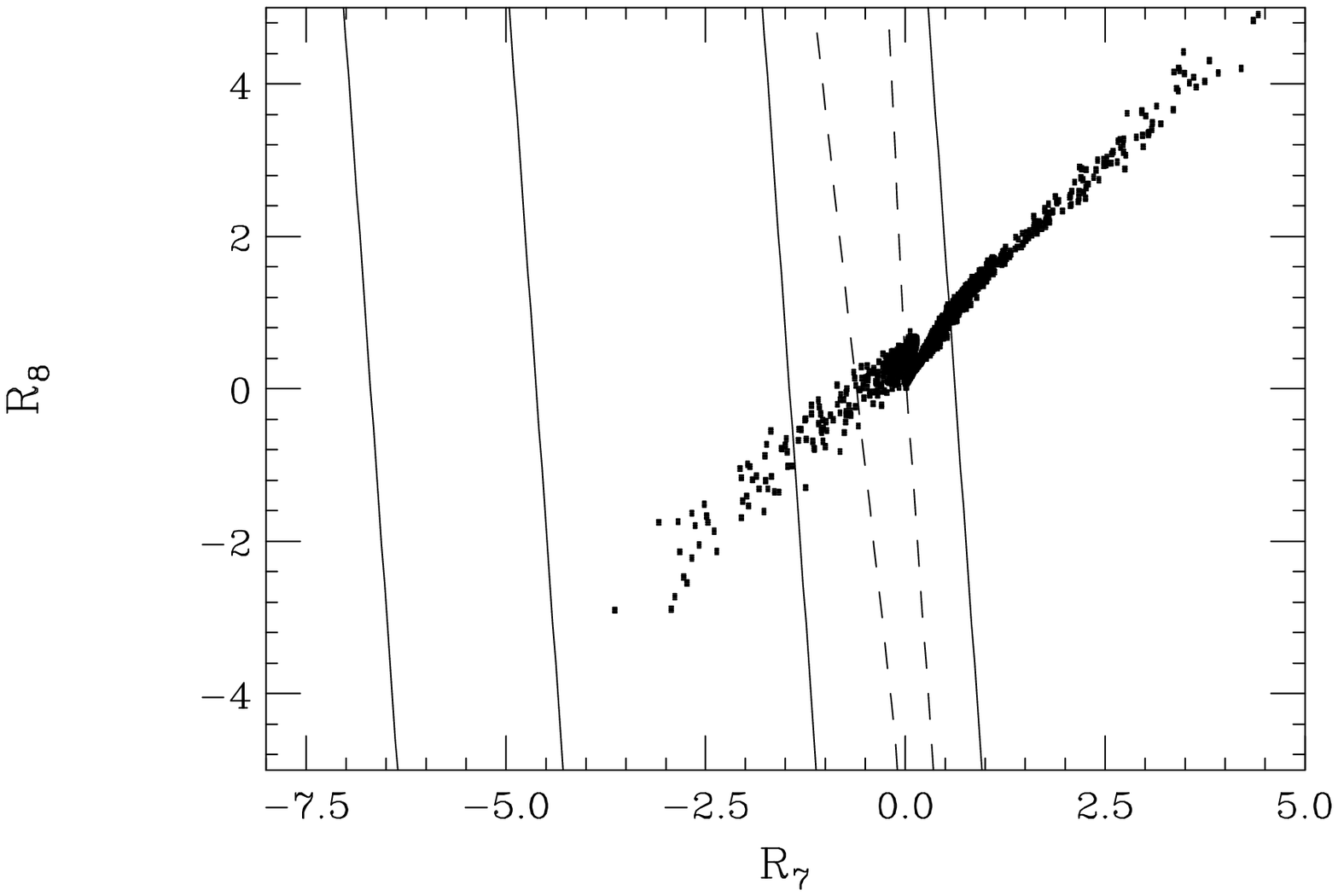,height=2.0in,width=2.20in}
\hspace*{5mm}
\psfig{figure=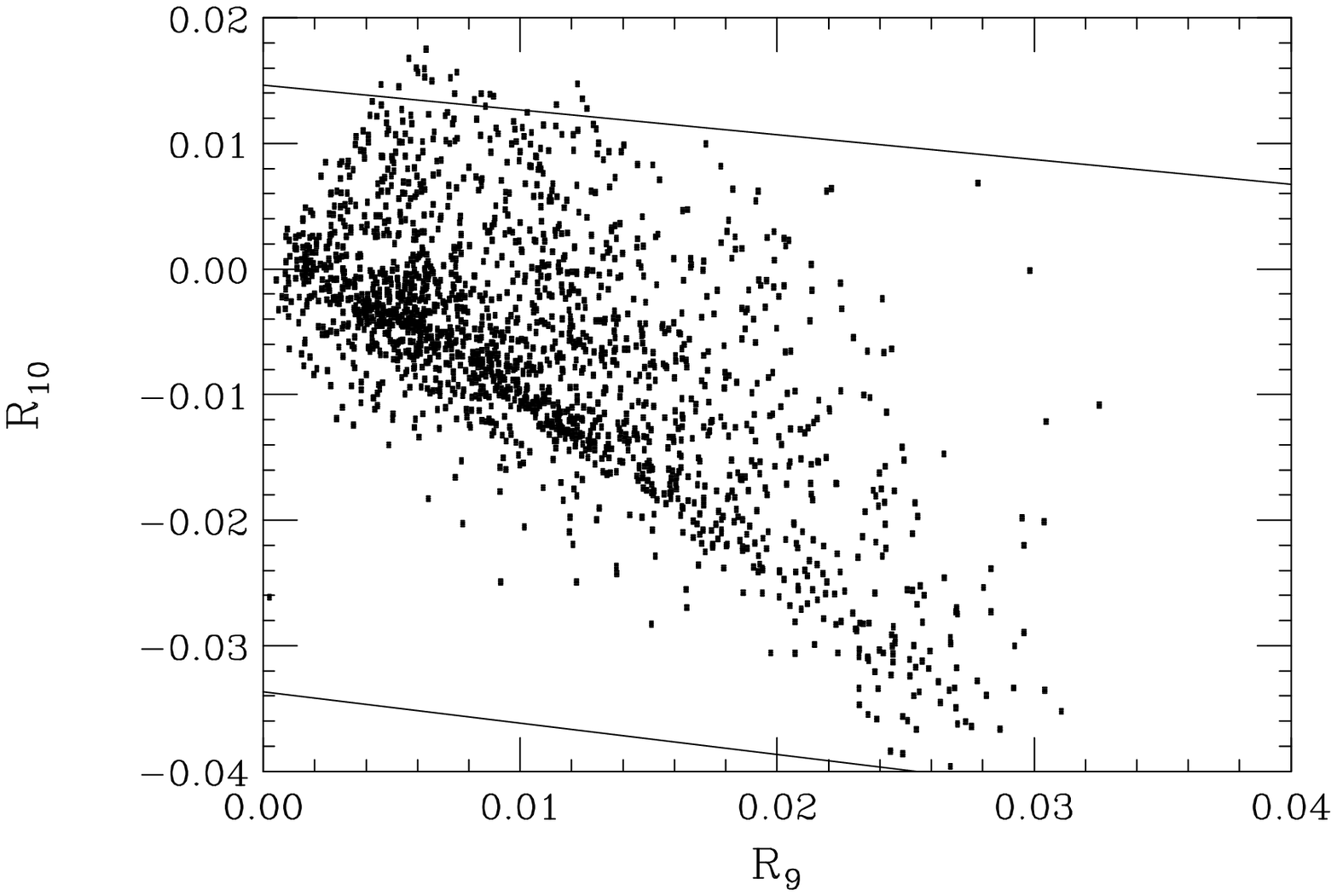,height=2.0in,width=2.20in}}
\vspace*{0.10cm}
\caption{(a) Parameter space 
scatter plot of $R_7$ vs. $R_8$ in the minimal supergravity model.
The allowed region from CLEO data lies inside the
solid diagonal bands.  The dashed band represents the potential future
$10\%$ measurement of \bsg\ as described in the text.
(b) Parameter space scatter plot of $R_9$ 
vs. $R_{10}$. The global fit to the coefficients
obtained with $5\times 10^8\, B\bar B$ pairs
corresponds to the region inside the diagonal bands.}
\label{fig4}
\end{figure}

We analyze the supersymmetric contributions
to the Wilson coefficients  \cite{jimjo,rareb} in terms of the quantities
\beq
R_i\equiv \frac{C^{susy}_i(M_W)}{C^{SM}_i(M_W)}-1\equiv 
{C_i^{new}(M_W)\over C_i^{SM}(M_W)}\,,
\eeq
where $C^{susy}_i(M_W)$ includes the full SM plus superpartner
contributions.  $R_i$ is meant to indicate the fractional deviation
from the SM value.  We will search over the full parameter space of the 
minimal supergravity model, calculate the $R_i$ for each generated point in
the supersymmetric parameter space, and then compare with the expected
ability of $B$ Factories to measure the $R_i$ as determined by our global
fit to the Wilson coefficients.  We generate \cite{kkrw} these supergravity 
models by applying common soft scalar 
and common gaugino masses at the boundary scale.
The tri-scalar $A$ terms
are also input at the high scale and are universal.
The radiative electroweak symmetry 
breaking conditions yield the $B$
and $\mu^2$ terms as output, with a $\mbox{sign}(\mu )$ ambiguity left over.
(Here $\mu$ refers to the Higgsino mixing
parameter.) We also choose $\tan\beta$ and restrict it
to a range which will yield perturbative Yukawa couplings up to the GUT scale.
We have generated thousands of solutions according to the above procedure.  The
ranges of our input parameters are $0< m_0 < 500\gev$, $50< m_{1/2} < 250\gev$,
$-3 < A_0/m_0 < 3$, $2 < \tan\beta < 50$,
and we have taken $m_t=175\gev$.  Each supersymmetric solution is
kept only if it is not in violation of present constraints from
SLC/LEP and Tevatron direct sparticle production limits, and it is out of reach
of LEP II.  For each of these remaining solutions 
we now calculate $R_{7-10}$.
Our results are shown in the scatter plots of Fig.~\ref{fig4}
in the (a) $R_7-R_8$  and (b) $R_9-R_{10}$ planes.  The diagonal
bands represent the bounds on the Wilson coefficients as previously
determined from our global fit.  We see from Fig. 3(a) that the current CLEO 
data on \bsg\
already place signigicant restrictions on the supersymmetric parameter space,
whereas the minimal supergravity contributions to
$R_{9,10}$ are predicted to be essentially unobservable.  

\begin{figure}[t]
\centerline{
\psfig{figure=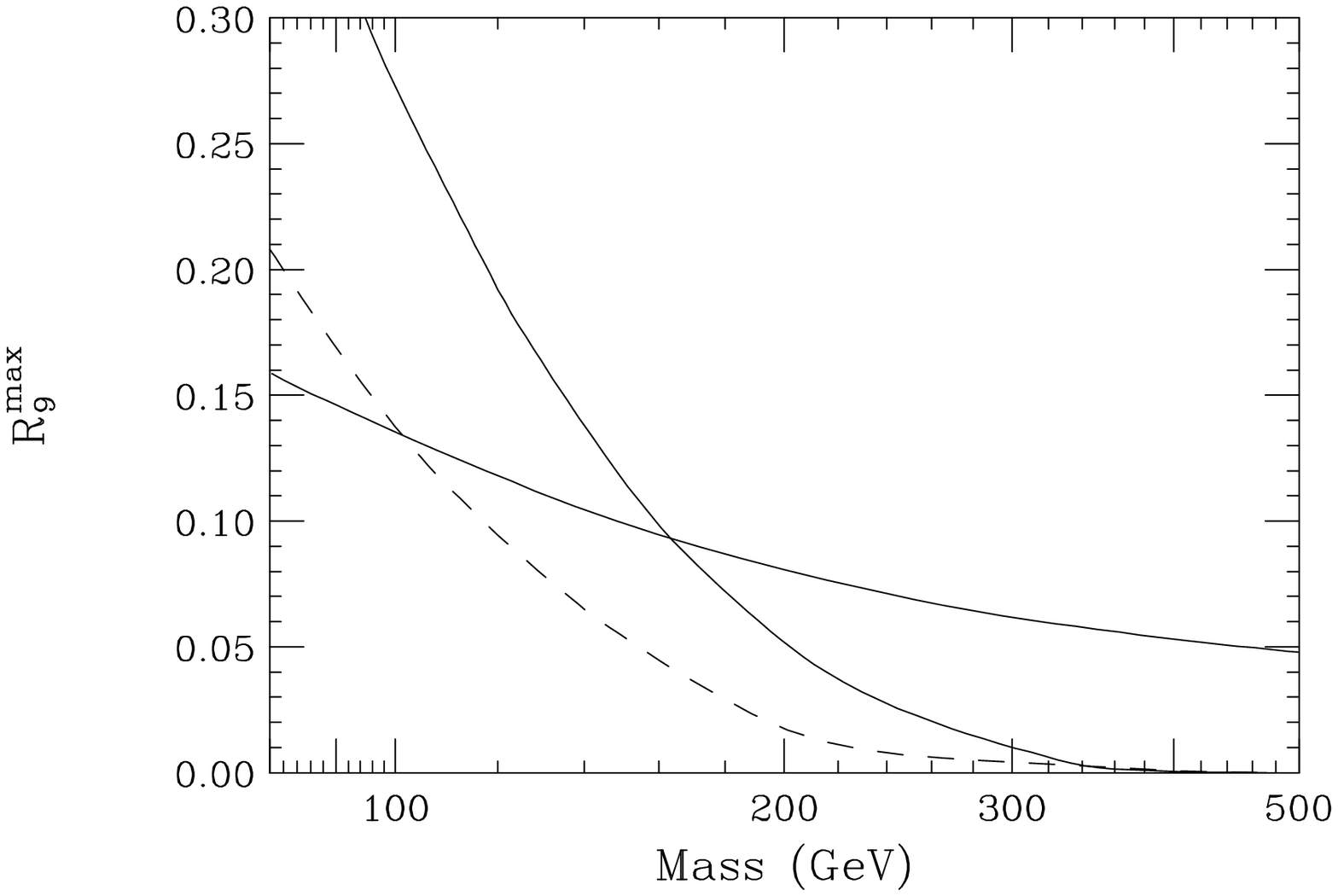,height=2.0in,width=2.20in}
\hspace*{5mm}
\psfig{figure=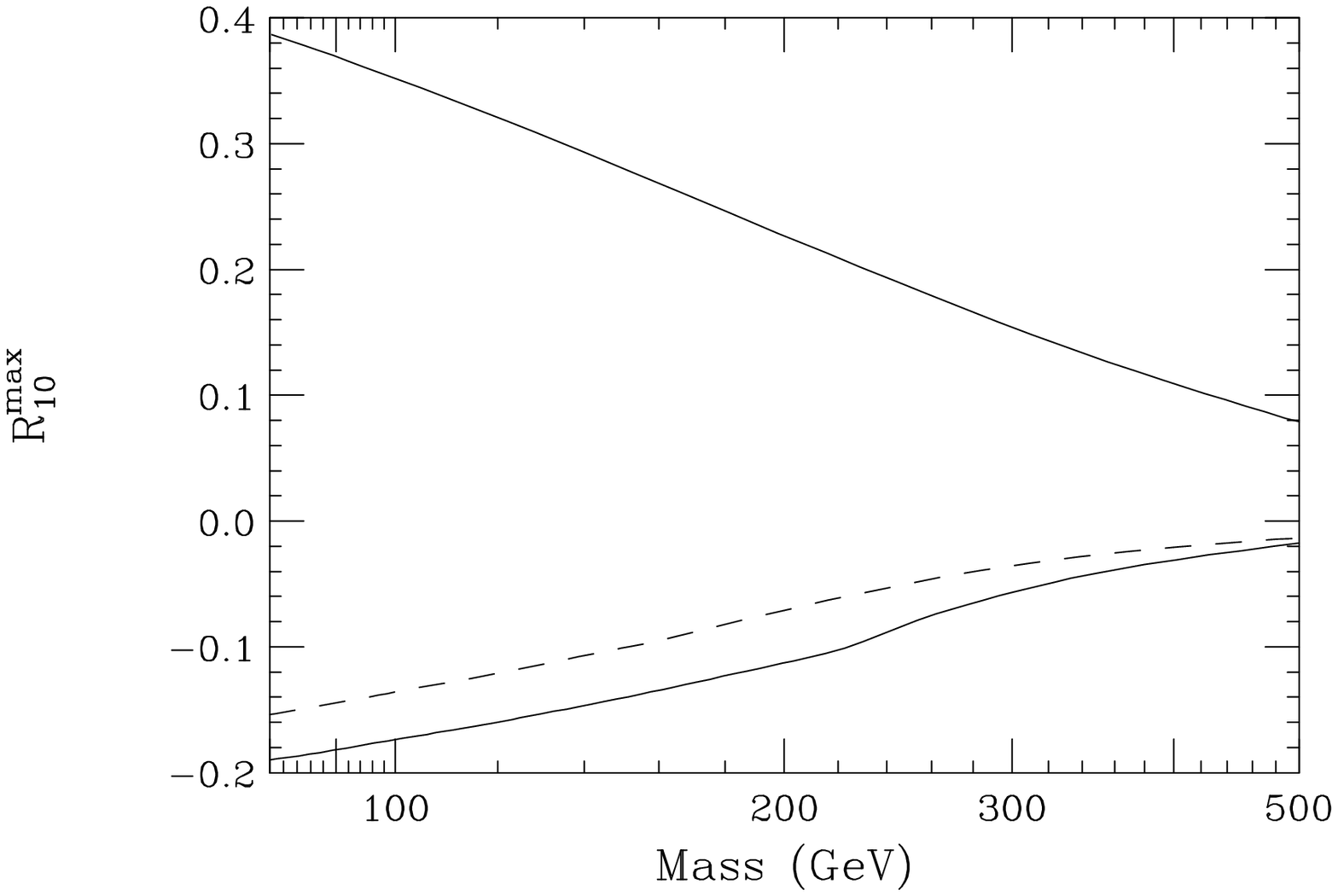,height=2.0in,width=2.20in}}
\vspace*{0.10cm}
\caption{The maximum value of (a) $R_9$ and (b) $R_{10}$ achievable for general 
supersymmetric models. The top solid line comes from the $t-H^\pm$
contribution and is displayed versus the $H^\pm$ mass.  The bottom solid line
is from the $\tilde t_i -\chi^\pm_j$ contribution with $\tan\beta =1$
and is shown versus the $\chi^\pm_i$ mass.  The dashed line is
the $\tilde t_i -\chi^\pm_j$ contribution with $\tan\beta =2$.
The other mass parameters which are not plotted are chosen to be just above
the reach of LEP II and the Tevatron.}
\label{fig8}
\end{figure}

A second, more phenomenological approach is now adopted. 
The maximal effects for the parameters $R_i$ can be estimated
for a superparticle spectrum, independent of the high scale assumptions.
However, we still maintain the assumption that CKM angles alone constitute
the sole source of flavor violations in the full supersymmetric lagrangian.
We will focus on the region $\tan\beta \lsim 30$.
The most important features which result in large contributions are
a light $\tilde t_1$ state present in the SUSY spectrum and at least one
light chargino state.  For the dipole moment operators a light Higgsino state
is also important.  A pure higgsino and/or pure
gaugino state have less of an effect than two mixed states when searching
for maximal effects in $R_9$ and $R_{10}$ and 
we have found that $M_2 \simeq 2 \mu$ is optimal.
Fig.~\ref{fig8} displays the
maximum contribution to $R_{9,10}$
versus an applicable SUSY mass scale.  
The other sparticle masses which are not shown ($\tilde t_i$,
$\tilde l_L$, etc.) are chosen to be just above the reach of LEP II or
the Tevatron, whichever yields the best bound.  We see that the maximum 
size of $R_{9,10}$ is somewhat larger than what was allowed in 
the minimal supergravity model, due to the lifted restriction
on mass correlations.

\begin{figure}[t]
\centerline{
\psfig{figure=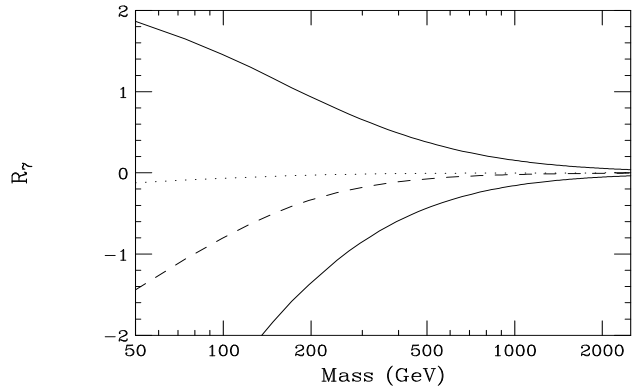,height=2.0in,width=3.25in}}
\vspace*{-0.20cm}
\caption{Contributions 
to $R_7$ in the different limits described in the text.  The top solid line
is the charged $H^\pm/t$ contribution versus $m_{H^\pm}$.  
The bottom solid line is the
$\tilde \chi^\pm_1/\tilde t_1$ contribution versus $m_{\tilde \chi^\pm}$
where both 
the chargino and stop are maximally mixed states with $\mu <0$.   
The dashed line is the $\tilde H^\pm /\tilde t_R$ contribution,
and the dotted line represents the $\tilde W^\pm /\tilde t_1$ contribution.
These two lines are both shown as a function of 
$\tilde \chi^\pm_1$ mass.
All curves are for $\tan\beta =2$ and $m_t=175\gev$.}
\label{fig11}
\end{figure}

Given the sensitivity of the 
observables it is instructive to narrow our focus to the coefficient of the
magnetic dipole operator.  The possibility exists that one
eigenvalue of the stop-squark mass matrix might be much
lighter than the other squarks, and 
we present results for $C_7(M_W)$ in the limit of
one light squark, namely the $\tilde t_1$, and light charginos.  
We allow the $\tilde t_1$ to have
arbitrary components of $\tilde t_L$ and $\tilde t_R$ since cross terms
can become very important.  This is especially noteworthy 
in the high $\tan\beta$ limit.  
We note that the total supersymmetric contribution to $C_7(\mw)$ 
will depend on several combinations of
mixing angles in both the stop and chargino mixing matrices and
cancellations can occur for different signs of $\mu$~\cite{garisto93:372}.  
The first case we examine is that where the lightest chargino is
a pure Higgsino and the lightest stop is purely right-handed:  
$\ca \sim \tilde H^\pm$, $\tilde t_1 \sim \tilde t_R$.  
The resulting contribution to $R_7$ is shown
as a function of the $\tilde t_R$ mass in Fig.~\ref{fig11} (dashed line) for
the case of $m_{\ca} \gsim M_W$.
Note that the SUSY contribution to $C_7(M_W)$ in this limit always adds
constructively to that of the SM.
Next we examine the limit where the light chargino is a pure Wino,
this contribution is shown in Fig.~\ref{fig11} (dotted line). 
The effects of a light pure Wino are small since (i) it couples with gauge 
strength rather than the top Yukawa, and (ii) 
supersymmetric models do not generally yield a light $\tilde t_L$ necessary
to couple with the Wino.  
Our third limiting case is that of a highly mixed $\tilde t_1$ state.
We find that in this case
large $\tan\beta$ solutions ($\tan\beta \gsim 40$)
can yield greater than ${\cal O}(1)$ contributions to
$R_7$ even for SUSY scales of $1\tev$! Low values of $\tan\beta$ can also
exhibit significant enhancements; this
is demonstrated for $\tan\beta =2$ in Fig.~\ref{fig11} (solid line).
We remark that large contributions are possible in this case
in both negative and positive directions of $R_7$ depending on the
sign of $\mu$.  We note that this is a region of SUSY parameter space
which is highly motivated by $SO(10)$ grand unified theories.

\begin{figure}[t]
\centerline{
\psfig{figure=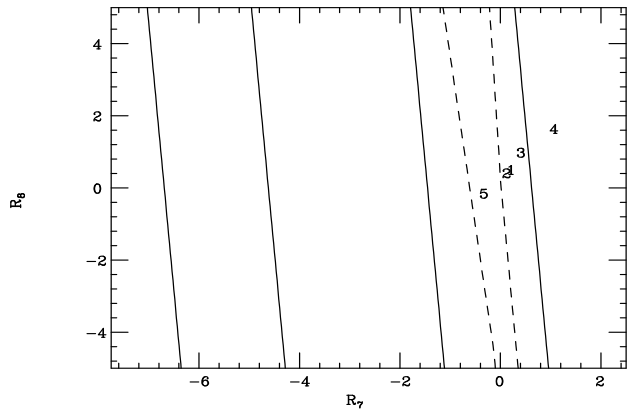,height=2.7in,width=3.95in,angle=-90}}
\vspace*{-0.50cm}
\caption{Values in the $R_7 - R_8$ plane for the five Snowmass NLC SUGRA points.
The solid and dashed bands represent the present bounds from CLEO and those
anticipated from future experiment, respectively, as described in Figure 3.}
\label{susysnow}
\end{figure}

Lastly, we compare the reach of rare $B$ decays in probing SUSY parameter
space with that of high energy colliders.  We examine a set of five points in
the minimal supergravity (SUGRA) parameter space that were chosen at Snowmass 
1996 \cite{snow} for the study of supersymmetry at the NLC.  Point $\# 3$ is
the so-called ``common'' point used for a comparison of SUSY studies at the 
NLC, LHC, and Tev33.  Once these points are chosen the sparticle
mass spectrum is obtained, as usual, via the SUGRA relations and their
contributions to \bsg\ can be readily computed.  The results are displayed
in the $R_7 - R_8$ plane in Fig.~\ref{susysnow}~(labeled $1-5$ for each SUGRA 
point), along with 
the bounds previously obtained from our fits to the present CLEO data and 
to anticipated future data assuming the SM is realized.
We see that four of the points should be discernable from the SM in
future measurements, and that one of the points is already excluded by CLEO!  

We thus conclude that rare $B$ decays are indeed complementary
to high energy colliders in searching for supersymmetry.

\section{Left-Right Symmetric Model in $b\to s$ Transitions}

The Left-Right Symmetric Model (LRM), which is based on the extended electroweak
gauge group $SU(2)_L\times SU(2)_R\times U(1)$ can lead to interesting new
effects in the $B$ system.\cite{lrmbc,lrmbsg}  Due to the extended gauge 
structure there are both new neutral and charged gauge bosons, $Z_R$ and $W_R$,
as well as a right-handed gauge coupling, $g_R$, which is subject to the
constraint $0.55<\kappa\equiv g_R/g_L<2.0$ from naturalness and GUT embedding
conditions.  The extended symmetry can be broken to the SM via the action of
Higgs fields that transform either as doublets or triplets under $SU(2)_R$.
This choice of Higgs representation determines both the mass relationship
between the $Z_R$ and the $W_R$ (analogous to the condition that $\rho=1$ in
the SM) as well as the nature of neutrino masses.  After complete symmetry
breaking the charged $W_R$ mixes with the SM $W_L$ to form the mass
eigenstates $W_{1,2}$ (where $W_1$ is the state which is directly produced at 
the Tevatron and LEP II).  This mixing is described by two parameters: a real
mixing angle $\phi$ and a phase $\alpha$.  In most models $\tan\phi$ is 
naturally of order of the ratio of masses $r=M_1^2/M_2^2$, or less,
in the limit of large $M_2$.  In this model the charged current interactions 
of the right-handed quarks are governed by a right-handed CKM matrix, $V_R$,
which, in principle, need not be related to its left-handed counterpart $V_L$.
$V_R$ will then involve 3 new angles as well as 6 additional phases, all of
which are {\it a priori} unknown parameters.  Phenomenological constraints
on the LRM are quite sensitive to variations of $V_R$.  If one assumes
manifest left-right symmetry, that is $V_R=V_L$ and $\kappa=1$, then
the $K_L-K_S$ mass difference implies that $M_R>1.6$ TeV.  However, if that
assumption is relaxed and $V_R$ (as well as $\kappa$) is allowed to vary
then $W_R$ masses as low as 500 GeV can be accomodated by present data.
This implies that the magnitude of $\tan\phi$ is $\leq{\mbox few}\times
10^{-2}$.

Clearly, this model contains many additional sources of CP violation; a
partial cataloging of the possible effects can be found in Ref. \cite{lrmcp}.
In addition, the influence of the LRM may be felt in both tree and loop-level
$B$ decays.  In particular, the possibility of a large right-handed component
in the hadronic current describing $b\to c$ transitions has long been a
subject of discussion\cite{lrmbc}.  Here we examine the possibility of using
the rare decays \bsg\ and \bsll\ as a new tool in exploring the parameter
space of the LRM.  The exchange of a $W_R$ within a penguin or box diagram,
in analogy with the SM $W_L$ exchange, can lead to significant deviations from
the SM predictions for the rates and kinematic distributions in these decays.

\begin{figure}[t]
\centerline{
\psfig{figure=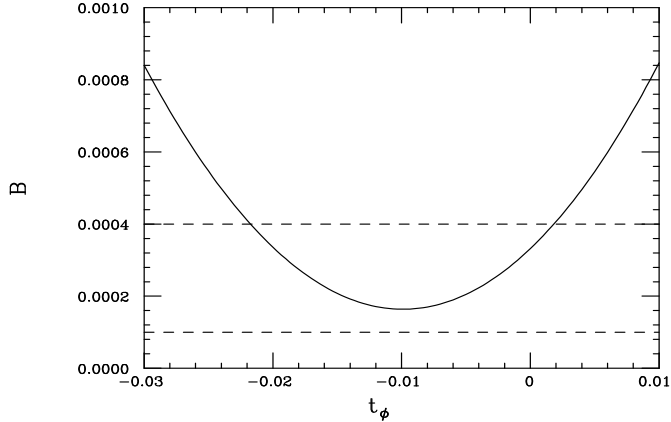,height=2.7in,width=3.95in,angle=-90}}
\vspace*{-0.50cm}
\caption{The \bsg\ branching fraction in the LRM for $m_t(m_t)=170$ GeV
as a function of $\tan\phi$, assuming $\kappa=1$, $V_R=V_L$, and
$M_2=1.6$ TeV.  The $95\%$ C.L. CLEO results lie inside the dashed lines.}
\label{lrmbsgfig}
\end{figure}

In the LRM the complete operator basis governing $b\to s$ transitions in
Eq. (\ref{effham}) must be expanded to 
\begin{equation}
{\cal H}_{eff}={-4G_F\over\sqrt 2} \sum_{i=1}^{12} C_{iL}(\mu){\cal O}_{iL}(\mu)
+L\to R \,.
\label{effhamlrm}
\end{equation}
This includes the right-handed counterparts to the usual 10 purely left-handed
operators, as well as two pairs of additional four-quark operators of mixed
chirality, 
${\cal O_{11L}}\sim (\bar s_\alpha c_\beta)_R(\bar c_\beta b_\alpha)_L$ and 
${\cal O_{12L}}\sim (\bar s_\alpha c_\alpha)_R(\bar c_\beta b_\beta)_L$.
The 2 subsets of left- and right-handed operators, 
${\cal O}_{1-12L,R}$ are decoupled and do not mix under REG evolution.  The
decay \bsg, where the operators ${\cal O}_{1-8,11,12(L,R)}$ contribute, has
been studied in some detail.\cite{lrmbsg}  In particular it was shown that
the left-right mixing terms associated with $\tan\phi\neq 0$ can be enhanced 
by a helicity flip factor of $\sim m_t/m_b$ and can lead to significantly
different predictions from the SM even when $V_L=V_R$ and $W_2$ is heavy.
This is depicted in Fig. \ref{lrmbsgfig} from Rizzo.\cite{lrmbsg}  It is also
clear from the figure that not only is the SM result is essentially obtained 
when $\tan\phi=0$, but also that a conspiratorial solution occurs when
$\tan\phi\simeq -0.02$.  From the \bsg\ perspective, these two cases are
indistinguishable, independent of any further improvements in the
measurement of the branching fraction.

\begin{figure}[t]
\centerline{
\psfig{figure=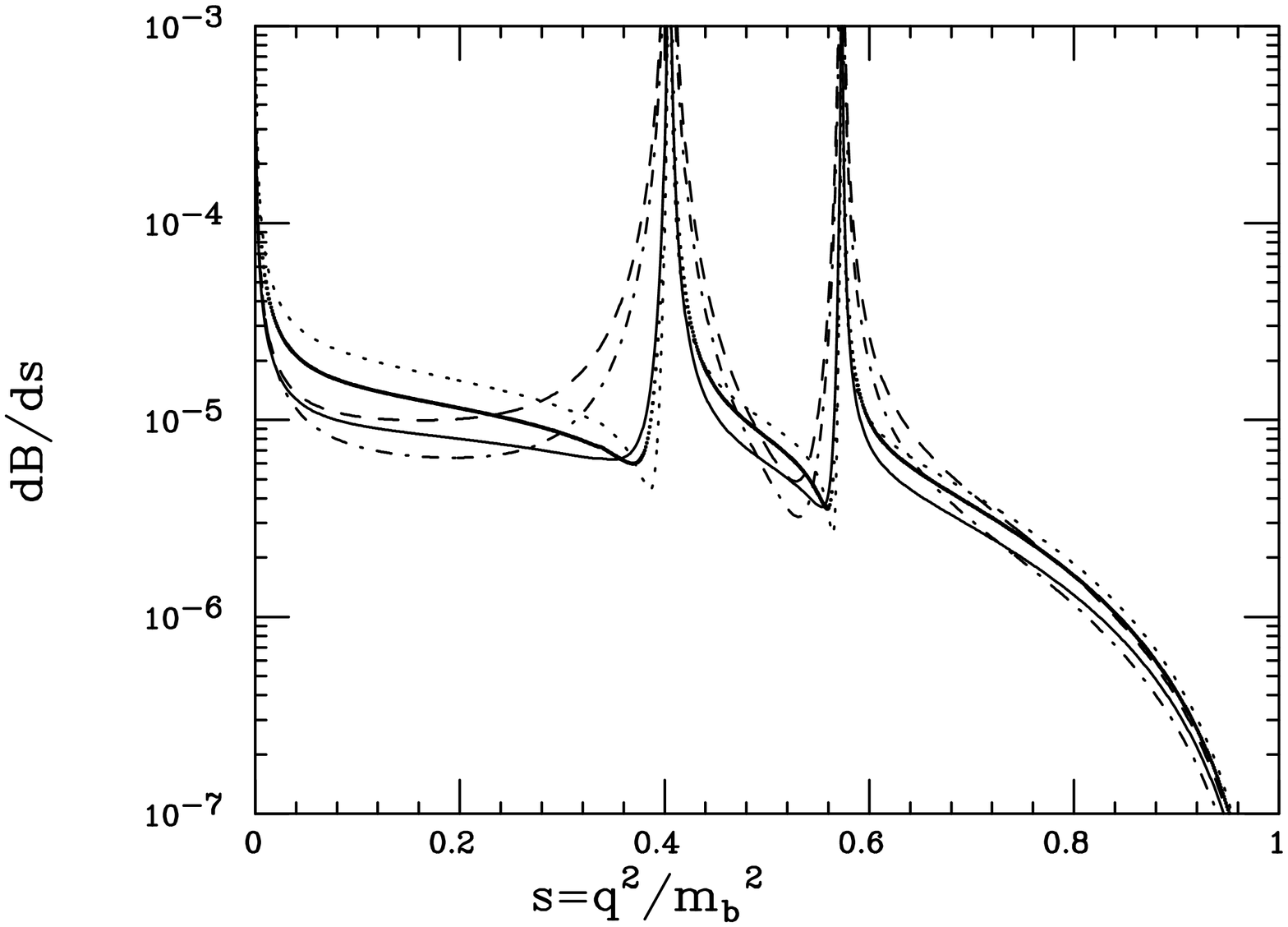,height=2.5in,width=2.250in,angle=-90}
\psfig{figure=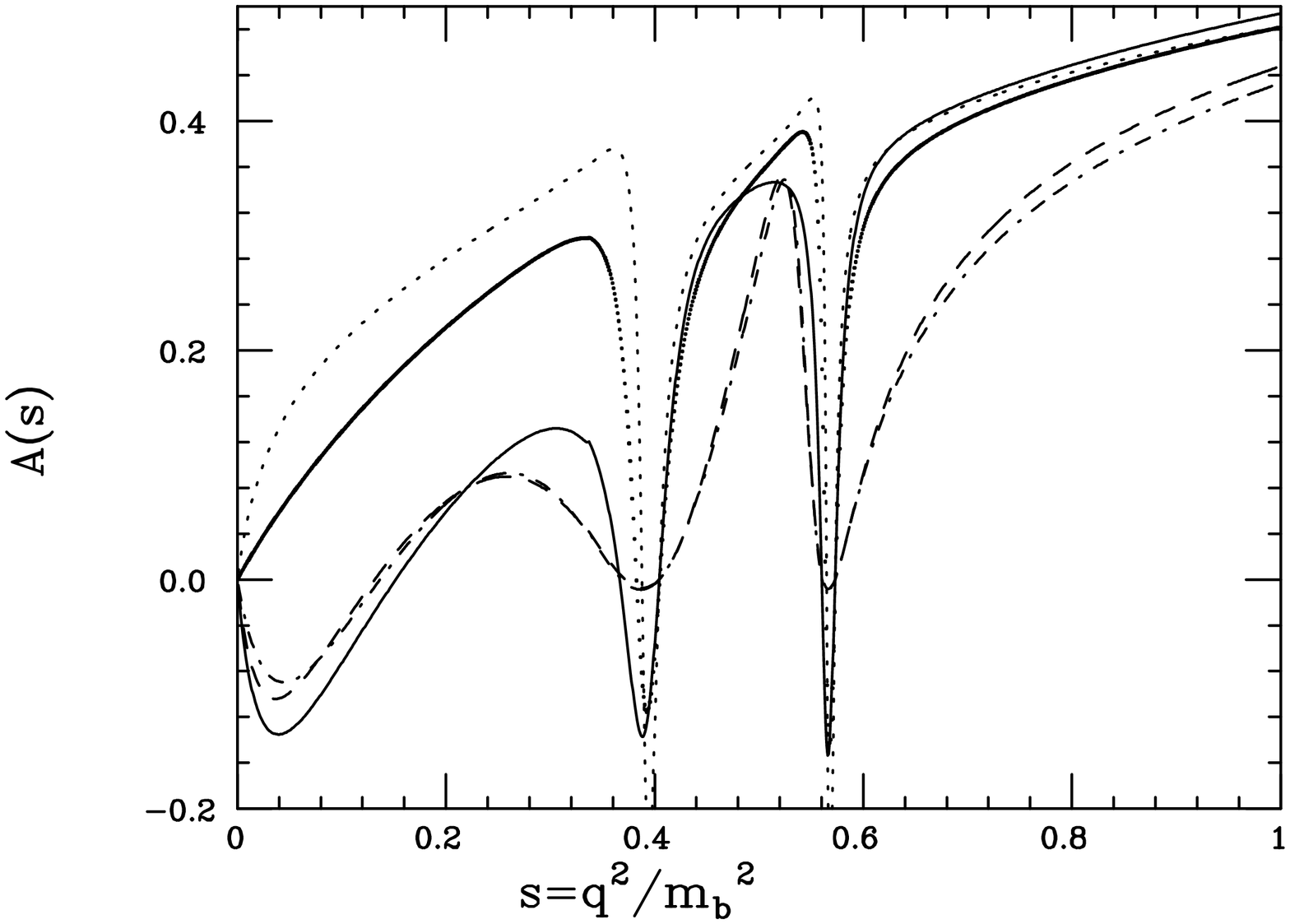,height=2.5in,width=2.250in,angle=-90}}
\vspace*{-0.50cm}
\caption{Differential decay distribution and lepton forward-backward
asymmetry for the decay \bsll\ in the SM (solid) and for four points in the
LRM parameter space which yield the SM value for $B(\bsg)$ and satisfy all
other existing experimental constraints.}
\label{bslllrm}
\end{figure}

LRM effects in \bsll\ have recently been examined by Rizzo,\cite{lrmtgr}
where it is found that the observables associated with this decay can
distinguish the LRM from the SM.  Here, all 24 operators in 
Eq. (\ref{effhamlrm}) participate in the renormalization and the matrix
element now depends on $C_{7,9,10L,R}(\mu\sim m_b)$.  The determination of
the matching conditions at the electroweak scale for these 24 operators is
tedious due to the large number of parameters, and in addition to new tree
graphs, 116 one-loop diagrams must be evaluated.  The predictions for the
lepton pair mass distribution and forward-backward asymmetry for 4 sample
points of the LRM parameter space is compared to the SM in Fig. \ref{bslllrm}.
These 4 sample points yield the same rate the decay \bsg\ as does the SM and
satisfy all other low-energy constraints and direct Tevatron searches.  As can
be seen from this figure, these sample LRM predictions not only differ from 
the SM, but also from each other.

The extension of the operator basis in Eq. (\ref{effhamlrm}) implies that
the conventional model independent determination of the Wilson coefficients
discussed above will not apply in this case.  In fact, this global fit
technique has recently been shown to fail\cite{lrmtgr} for the LRM, and in doing
so it provides a powerful probe for the existence of new operators.  In general,
there are
three ways new physics can affect the global fit to the coefficients: ($i$)
the numerical values for the coefficients are found to agree with SM
expectations with a good $\chi^2$; in this case the new physics is decoupled.
($ii$) A quality fit is obtained, but the fit values of the coefficients 
deviate from SM expectations. ($iii$) The $\chi^2$ value for the best three
parameter fit is found to be very large and cannot be explained by an
under estimation of systematic uncertainties.  It is this latter case which
indicates the existence of additional operators.  For the 4 LRM sample points 
discussed above, the 3 parameter ($C_{7,9,10L}$) global fit yields enormous 
values of $\chi^2/d.o.f.$, of order $1000/25$, for a data sample of 
$5\times 10^8$ $B\bar B$ pairs, clearly signaling an inconsistent fit.
For smaller data samples, \ie, $5\times 10^7$ $B\bar B$ pairs, the results
are more  dependent on the particular values of the LRM parameters.  However,
with sufficient statistics, it will be possible to observe the case where the
canonical three coefficient fit fails, revealing not only the existence of
physics beyond the SM, but that this new physics requires an extension of
the operator basis.

\section{Conclusions}

This talk focused on supersymmetric and left-right symmetric model effects,
as well as model independent tests for new physics, in rare $B$ decays.
Of course, there are numerous other candidates for physics models beyond
the SM, as well as many other reactions where they can be tested.  A brief
compendium of these is given in Table \ref{newphystab}.  Here, we display the
effects of ($i$) Multi-Higgs-Doublet Models (MHDM), with and without Natural 
Flavor Conservation (NFC), ($ii$) the Minimal Supersymmetric Standard Model
(MSSM), and the supersymmetric models with squark alignment, effective SUSY
scales, and R-parity violation, ($iii$) the LRM, with and without manifest
left-right symmetry in the quark mixing matrices, ($iv$) a fourth generation,
and ($v$) models with $Z$-boson mediated FCNC.  We describe whether these
models have the potential to cause large deviations from SM predictions in 
rare decays and $D^0-\bar D^0$ mixing, whether new phases exist which
contribute to $B_D^0-\bar B_d^0$ mixing, and whether the new physics effects
cancel in the ratio of mass differences in the $B_s$ to $B_d$ systems.  This
table is only intended to give a quick indication of potential effects.

In conclusion, we see that the $B$ sector can provide a powerful probe,
not only for the existence, but also for the structure of physics beyond the 
SM.

\begin{table}
\centering
\begin{tabular}{|c|c|c|c|c|}\hline\hline
Model & Rare Decays & $\Delta M_s\over\Delta M_d$ & New Phase 
& $D^0-\bar D^0$ \\
&  &  & $B_d$ Mixing & Mixing \\ \hline
MHDM: with NFC& \bsg\ & $=$ SM & No & Small\\
~~~~: no NFC & Not Really & $\neq$ SM & Yes & Big\\ \hline
MSSM & \bsg\ & $=$ SM & No & Small\\
Squark Alignment & Small & $\sim$ SM & No & Huge\\
Effective SUSY & Small & $\sim$ SM & Yes & Small\\
R-Parity Violation & Big & $\neq$ SM & Yes & Big\\ \hline
LRM: $V_L=V_R$ & \bsg\ and & $=$ SM & No & Small \\
~~~: $V_L\neq V_R$ & \bsll\ & $\neq$ SM & Yes & Big \\ \hline
4th Generation & Big & $\neq$ SM & Yes & Huge \\
$Z$-mediated FCNC & Big & $\neq$ SM & Yes & Big \\ \hline\hline
\end{tabular}
\caption{Model dependent effects of new physics in various processes.}
\label{newphystab}
\end{table}

\end{document}